# Observation of emergent Dirac physics at the surfaces of acoustic higher-order topological insulators

Fei Meng[1,2,#], Zhi-Kang Lin[3,#], Weibai Li[2], Peiguang Yan[4], Yun Zheng[5], Jian-Hua Jiang[3,*], Baohua Jia[2,*], Xiaodong Huang[2,*]

1. Hubei Key Laboratory of Roadway Bridge & Structure Engineering, Wuhan University of Technology, Wuhan, Hubei 430070, China
2. Centre of Translational Atomaterials, Faculty of Science, Engineering and Technology, Swinburne University of Technology, Hawthorn VIC 3122, Australia
3. School of Physical Science and Technology, and Collaborative Innovation Center of Suzhou Nano Science and Technology, Soochow University, Suzhou 215006, China.
4. Key Laboratory of Optoelectronic Devices and Systems, College of Physics and Optoelectronic Engineering, Shenzhen University, Shenzhen 518060, China
5. State Key Laboratory of Geomechanics and Geotechnical Engineering, Institute of Rock and Soil Mechanics, Chinese Academy of Sciences, Wuhan 430071, China

#These authors contributed equally.
∗ Corresponding authors:
Jian-Hua Jiang: jianhuajiang@suda.edu.cn
Baohua Jia: bjia@swin.edu.au
Xiaodong Huang: xhuang@swin.edu.au

**Abstract**: Using three-dimensional (3D) sonic crystals as acoustic higher-order topological insulators (HOTIs), we discover two-dimensional (2D) surface states described by spin-1 Dirac equations at the interfaces between the two sonic crystals with distinct topology but the same crystalline symmetry. We find that the Dirac mass can be tuned by the geometry of the two sonic crystals. The sign reversal of the Dirac mass reveals a surface topological transition where the surface states exhibit zero refractive index behavior. When the surface states are gapped, one-dimensional (1D) hinge states emerge due to the topology of the gapped surface states. We confirm experimentally the zero refractive index behavior and the emergent topological hinge states. Our study reveals a multidimensional Wannier orbital control that leads to extraordinary properties of surface states and unveils an interesting topological mechanism for the control of surface waves.

## 1. Introduction

Topological insulators (TIs) are intriguing materials which behave as insulators in the bulk but as conductors on the edges [1,2]. For instance, two-dimensional (2D) TIs host topologically protected one-dimensional (1D) edge states, while 3D TIs host 2D topological surface states. In the past years, higher-order topological insulators (HOTIs) extend the conventional bulk-edge correspondence to higher-order bulk-boundary correspondences, leading to rich multidimensional topological phenomena [3-9]. For instance, a *d*-dimensional HOTI can host (*d*-1)- and (*d*-2)-dimensional topological boundary states simultaneously. Recently, the concept of HOTI is generalized to classical systems such as mechanical metamaterials [10-14], electrical circuits [15-24], sonic crystals [25-35], and photonic crystals [36-49]. Because of their macroscopic controllability and the convenience in the excitation and detection of acoustic waves, sonic crystals stand out as a versatile platform for higher-order topological physics [18-23,25-27]. By designing the solid or fluid scatters/resonators in sonic crystals, we can conveniently form the desired energy bands, and the entire spectrum is easily accessible comparing to electronic systems. This allows us to realize some special manipulation of sound waves in acoustic HOTIs.

Here, we report on the experimental discovery of surface topological transitions and the emergent spin-1 Dirac physics at the interfaces between two 3D acoustic HOTIs, which are triggered by the tunable higher-order topology through the geometry control of the sonic crystals. Two acoustic HOTIs are created by using sonic crystals of the same crystalline symmetry but distinct higher-order topology due to different geometry. We find that the 2D boundary states emerging at the interfaces between the two acoustic HOTIs can be described by the spin-1 Dirac equation. The emergent surface Dirac physics is determined by the configurations of the Wannier orbitals in the two HOTIs. By tuning the geometry of two acoustic HOTIs, the Dirac mass of the 2D surface states can be tuned to experience a topological transition (i.e., a sign reversal of the Dirac mass). At the transition point, the surface states become 2D massless Dirac waves in the bulk band gap, which exhibit extraordinary properties such as zero refractive index behavior. As the 2D Dirac mass becomes finite, topological hinge states emerge due to the topology of the gapped surface states. We experimentally confirm the zero refractive index behavior and 1D sound wave propagation via topological hinge states. The findings demonstrate the rich physics and phenomena in higher-order topological materials.

## 2. 3D acoustic HOTIs

We design two sonic crystals (SC1 and SC2) of the same spatial symmetry to realize different topological properties. The unit cell of SC1 has six acoustic resonators on the faces of a cubic by coupling them via the overlapped air regions. Then SC2 is realized through shifting the unit cell of SC1 by the vector (0.5$a$, 0.5$a$, 0.5$a$). $a$=16 mm is the lattice constant. Each acoustic resonator consists of two overlapping pyramid air regions, as shown in Fig. 1a. In this work, the geometric parameters of the pyramids $b_1$ and $b_2$ are taken as $0.955a$ and $0.675a$, respectively, while $h_1$ and $h_2$ varies. Fig. 1b illustrates the air regions in SC1 and SC2 for $h_1 = h_2 = 0.175a$ (see details in Supplemental Material Section 1). The sonic crystal structures are fabricated by photosensitive resin using the 3D-printing technology. SC1 and SC2 exhibit the simple cubic lattice symmetry, i.e., the space group $P_{m\bar{3}m}$.

With such constructions, SC1 and SC2 have the same acoustic band dispersion (Fig. 1c). However, the symmetry properties of acoustic bands are different. In particular, the parity eigenvalues of the Bloch states at the high symmetry point X are different for SC1 and SC2, which leads to distinct topology. The difference is first characterized by the total bulk polarization of the bands below the first band gap, $\mathbf{P} = (P_x, P_y, P_z)$. The symmetry of the simple cubic lattice guarantees $P_x = P_y = P_z$ and quantizes $P_i$ ($i = x, y, z$) to either 0 or 1/2. We find that $\mathbf{P} = (0, 0, 0)$ for SC2, whereas $\mathbf{P} = (1/2, 1/2, 1/2)$ for SC1 (see details in Supplemental Material Section 2) [22]. To fully understand the higher-order topology, it is necessary to explore the Wannier orbitals in both acoustic HOTIs. We find that the Wannier orbitals are *s*-like orbitals, yet they locate at different positions for the two sonic crystals. For SC1, the

Wannier centers locates at the six surface centers of the unit cell. For SC2, the Wannier centers locate at the twelve hinge centers of the unit cell (see Supplementary Material Section 3).

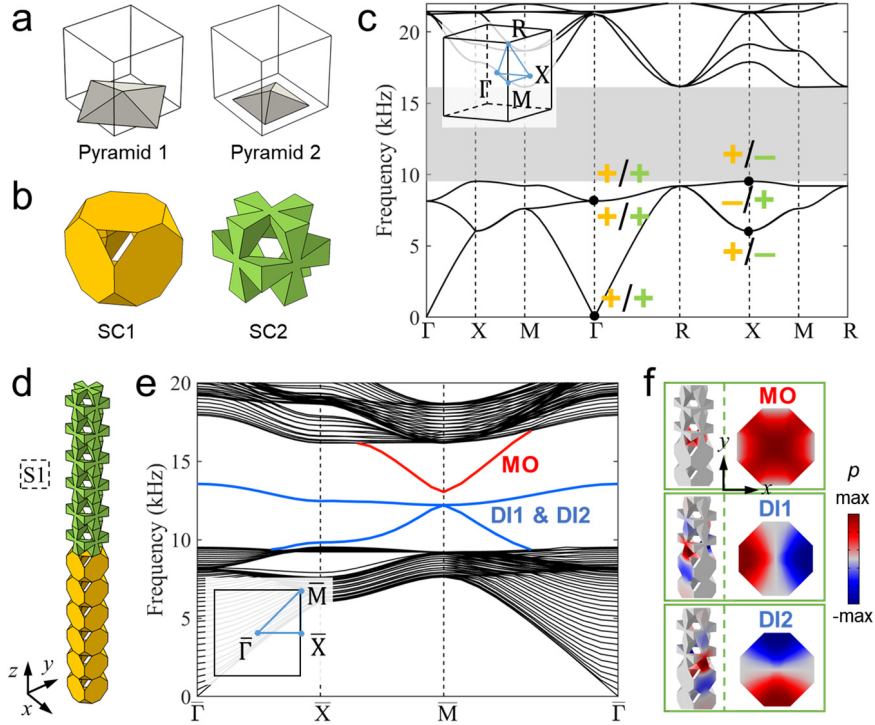

Fig. 1 | (Color online) 3D sonic crystals and topological surface states. (a) Illustrations of the unit cell structures for SC1 and SC2 when $h_1 = h_2 = 0.175a$. (b) Identical acoustic band structures of SC1 and SC2. A complete band gap appears from 9.53 kHz to 16.13 kHz. Here, + (–) denotes even (odd) parity of the acoustic Bloch states at the Γ and X points (Their wavefunctions are shown in Supplemental Material Section 2). Yellow (green) symbols are for SC1 (SC2). (c) The supercell S1 consists of SC1 and SC2. It is finite in the $z$ direction but periodic in the $x$ and $y$ directions. (d) 2D acoustic band structure of S1. Three surface bands emerge in the bulk band gap. They are labeled as DI1, DI2, and MO according to their parity properties at the $\bar{M}$ point of the surface Brillouin zone (see the inset). (e) Acoustic wavefunctions (i.e., acoustic pressure, $p$, profiles) of the surface states at the $\bar{M}$ point. Left: 3D view. Right: wavefunctions on the interface between SC1 and SC2.

To study the 2D topological surface states, we construct a ribbon-like supercell shown in Fig. 1d. The supercell (denoted as S1) is finite in the $z$ direction, but periodic along the $x$ and $y$ directions. The calculated acoustic band structure for S1 is presented in Fig. 1e. In the bulk band gap, three surface bands emerge, which can be described by the spin-1 Dirac equation around the $\bar{M}$ point of the surface Brillouin zone, as shown in Fig. 1f.

The emergence of spin-1 Dirac physics at the interfaces between SC1 and SC2 is triggered by the topological properties of the two sonic crystals. According to topological theory [2], the

Wannier centers exposed at the interface form the surface bands. There are in total three such Wannier centers located at the center and the two edge centers of the interface unit cell (see Supplementary Material Section 3). The former comes from SC1, while the latter two come from SC2. These Wannier centers form an emergent Lieb lattice at the interface leading to spin-1 Dirac physics at the $\bar{M}$ point. We find that the Dirac mass is controlled by the energy of these Wannier orbitals. When they have the same energy, a massless spin-1 Dirac cone emerges. In other situations, the spin-1 Dirac mass is finite, and the surface states are gapped (see Supplementary Material Section 4).

The symmetry properties of the surface bands can be reflected by acoustic pressure fields at the $\bar{M}$ point, which is shown in the insets of Fig. 1f. The acoustic wavefunctions indicate that there are one $s$-like mode (labeled as MO) and two $p$-like modes (labeled as DI1 and DI2) at the $\bar{M}$ point, which are consistent with the interface symmetry. Exploiting a $\boldsymbol{k} \cdot \boldsymbol{p}$ theory in the basis of the MO, DI1, and DI2 states, we find that the effective Hamiltonian for phonons around the $\bar{M}$ point is given by

$$H(\boldsymbol{q}) = \begin{bmatrix} 0 & iq_x v & iq_y v \\ -iq_x v & m & 0 \\ -iq_y v & 0 & m \end{bmatrix} + f_0, \tag{1}$$

where $m$ and $v$ are the Dirac mass and velocity, respectively. $f_0$ is the frequency of the Dirac point. The above analysis is also supported by an effective 2D theory based on a tight-binding Lieb lattice model (see Supplemental Material Section 5), confirming the intriguing emergent Dirac physics at the 2D interfaces.

Calculations show that the frequencies of MO and DI states can be tuned by controlling the geometry of SC1 and SC2. The results are presented in Fig. 2a where we choose to keep $h_1$ fixed and change $h_2$ for either SC1 or SC2. Such tuning can flip the frequency order of the MO and DI states at the $\bar{M}$ point. The results indicate that the Dirac mass can be tuned to undergo a sign reversal, leading to a surface topological transition. We find that the positive Dirac mass phase is trivial and has a Wannier center at the center of the interface unit cell. The negative Dirac mass phase is topological and has two Wannier centers at the edge centers of the interface unit cell. The above picture reveals the underlying principles of the emergent Dirac physics at the surface of the HOTIs and opens a pathway for engineering the topological interface states.

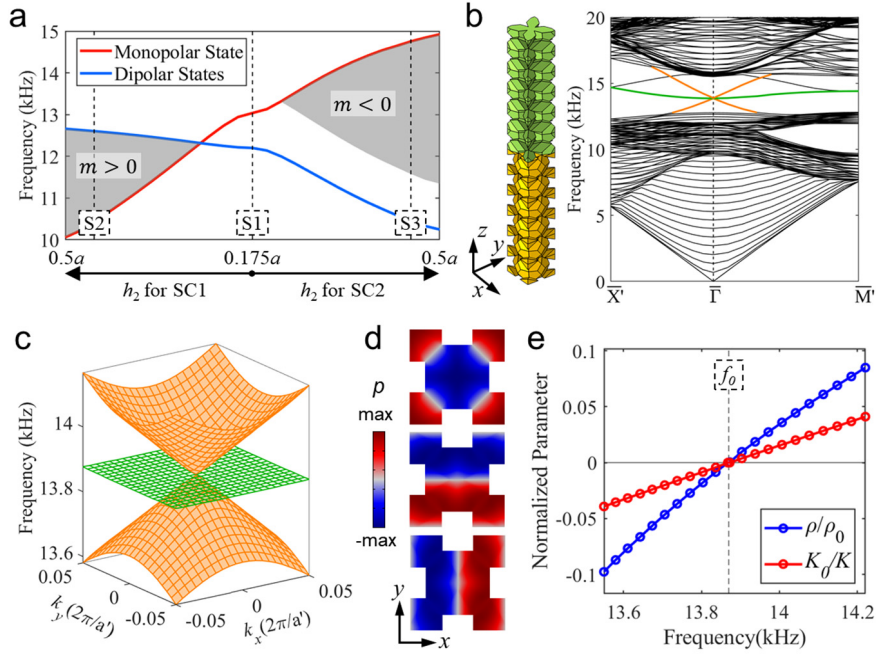

Fig. 2 | (Color online) Surface topological transition. (a) Frequencies of monopolar and dipolar states at the $\bar{M}$ point as a function of $h_2$. Note that we consider the change of $h_2$ in either SC1 (left) or SC2 (right), while other parameters remain unchanged. Gray areas denote the regions with a complete acoustic band gap for the surface states. The closing and reopening of the surface band gap accompany the sign change of the Dirac mass and the parity inversion of the surface bands at $h_2 = 0.265a$ for SC1. (b) The expanded supercell (left) and its band diagram (right), leading to a Dirac cone at the $\bar{\Gamma}$ point associated with the surface states in the bulk band gap. (c) Conical dispersion in the vicinity of the Dirac point. $a' = \sqrt{2}a$ is the lattice constant of expanded supercell in the *x-y* plane. (d) Acoustic wavefunctions of the surface states on the interface plane between SC1 and SC2. (e) The effective mass density and effective inverse bulk modulus. We use the density $\rho_0$ and bulk modulus $K_0$ of air to normalize the effective parameters for better graphic presentations ($\rho_0$ = 1.21 kg·m$^{-3}$, $K_0 = 1.42 \times 10^5$ Pa). Dashed line indicates the frequency of the Dirac point, $f_0$=13.87 kHz.

## 3. Zero-index conical surface states

When the Dirac mass of the surface states vanishes, a conical dispersion emerges in the surface states at the corner of the surface Brillouin zone (i.e., the $\bar{M}$ point). To trigger the zero refractive index behavior, we employ the Brillouin zone folding technique (see Supplemental Material Section 6) to bring the conical dispersion to the center of the Brillouin zone [34]. This can be realized by doubling the interface unit cell size and using an expanded supercell with a new set of parameters, i.e., $h_1 = h_2 = 0.27a$ for SC1 and SC2. With such parameters, we create gapless surface acoustic bands (see Supplementary Material Section 7) with a spin-1 Dirac cone at the surface Brillouin zone center (see Figs. 2b-d). The Dirac point has a frequency of 13.87 kHz, which is near the middle of the bulk band gap. Protected by the bulk gap, such a surface Dirac cone with a clean dispersion creates an unprecedented realm for the manipulation of surface acoustic waves, as shown below.

Using the effective media theory [50-53], we calculate the effective mass density $\rho$ and the effective inverse bulk modulus $1/K$ (i.e. the compressibility) for the surface acoustic waves near the Dirac point [35,36] (see details in Supplemental Material Section 8). We find that these quantities undergo simultaneous sign changes at the Dirac point (Fig. 2e). The simultaneous zero effective mass density $\rho$ and zero effective inverse bulk modulus $1/K$ at the surface Dirac point indicate the zero refractive index property [50-53]. This property can be used to collimate the surface acoustic waves emitting from a straight edge of the interface. Our experimental setup is illustrated in Fig. 3(a). The sample is a block with 16×5 expanded supercells [i.e., the supercell illustrated in Fig. 2(b)]. The left surface of the sample is glued by an acrylic slab to seal the airborne sound waves. We drill a small hole with a diameter of 5 mm in the center of the slab. The acoustic wave from the source is guided into the system via a plastic tube connected to the hole. For acoustic waves with a wavelength larger than 5 mm, the hole can be regarded as a point-like source on the surface, which excites all possible surface states according to their frequencies. Near the surface Dirac point, such an excitation gradually merges into the excitation of the Dirac point state. The propagation of the state exhibits zero-index feature, which can be detected by the acoustic pressure profile at the surface on the other side of the sample [see Fig. 3(a)]. As shown in Fig. 3(b), when the excitation frequency is 14.00 kHz, being close to the frequency of the Dirac point ($f_0$=13.87 kHz), both the simulated and measured acoustic pressure and phase profiles show the planar wavefront parallel to the edge of the sample. This indicates excellent surface wave collimation, even a point source is used in the experiments. Furthermore, the consistency between the simulation and the measurement confirms the wave collimation due to the surface Dirac point. In contrast, when the excitation frequency is set to 12.00 kHz (i.e., being away from the Dirac point frequency), the simulated and detected wave patterns do not show this phenomenon, which is due to the coexistence of many surface states being excited simultaneously. Such a drastic contrast supports that the surface wave collimation is caused by the zero-index property of the surface Dirac point.

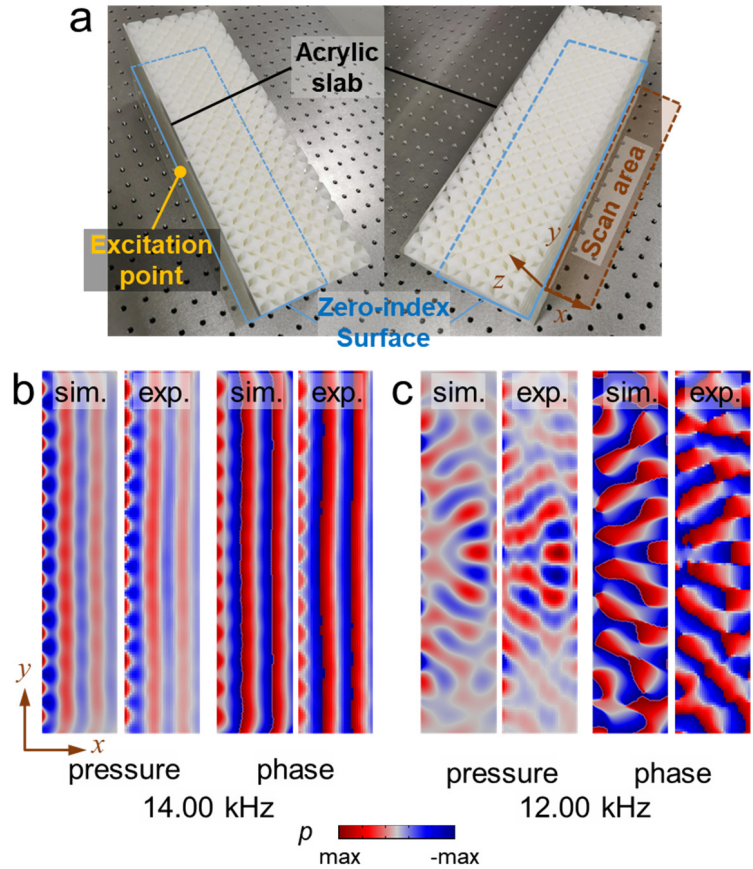

Fig. 3 | (Color online) Visualization of the double-zero-index property of the surface Dirac cone. (a) The fabricated sample and the experiment setup. The hole in the acrylic plate is used to guide acoustic waves into the system, while the detection is on the other side of the sample to detect the acoustic waves travel across the sample via the *x-y* interface in the middle. (b) and (c) The simulated and the measured acoustic pressure and phase profiles in the detection *x-y* plane when the excitation frequency is 14.00 kHz and 12.00 kHz, respectively.

## 4. 1D hinge states induced by surface topological band gaps

When the surface Dirac mass is finite, topological sound trapping at the 1D hinges can emerge due to the topology of the surface bands. As shown in Fig. 2(a), when the surface Dirac cone opens a band gap, it has either positive or negative Dirac mass. These two types of band gaps have distinct topology [54,55]. Protected by the four-fold rotation ($C_4$) symmetry of the interface, the topological surface band gap has a nontrivial topological index and Wannier centers at the edges of the interface unit cell, whereas the trivial surface band gap has the vanishing topological index and the Wannier center at the interface unit cell center. When a domain wall between such two interfaces is formed, 1D topological hinge states emerge due to the surface band topology, and it is robust against imperfections (Supplemental Material Section 9).

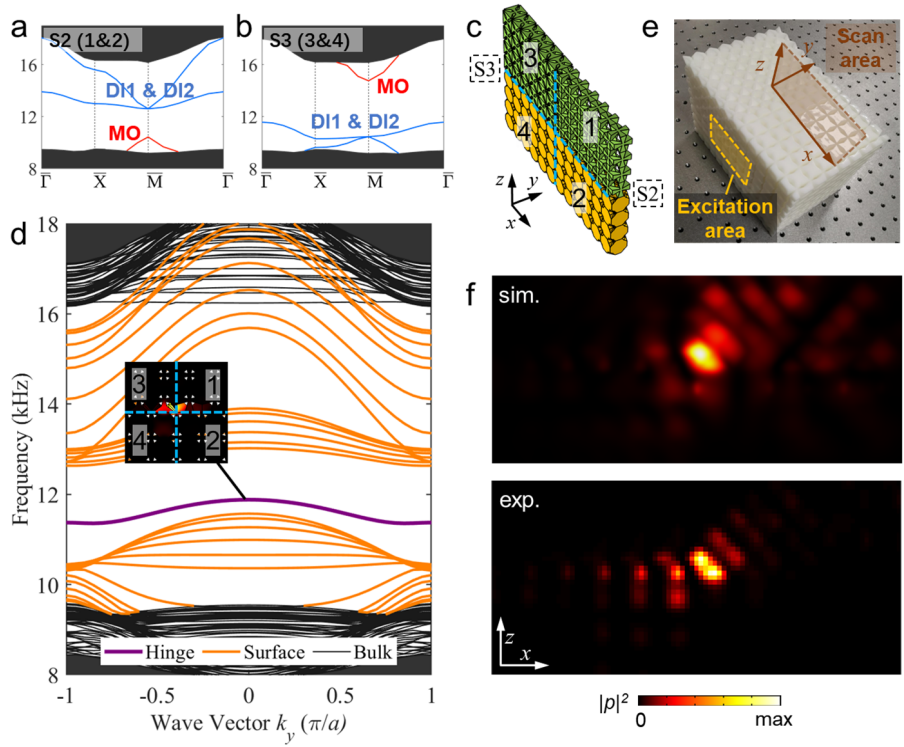

Fig. 4 | (Color online) 1D hinge state originates from inverted topological surface bands. (a) (b) Band diagrams of supercell S2 and S3 with gapped topological surface bands. (c) The slab-like hinge supercell consisted of S2 (right half) and S3 (left half). Zones 1 and 3 are the trivial sonic crystals in S2 and S3. Zones 3 and 4 are the non-trivial sonic crystals in S2 and S3. The boundaries between zone 1 and 2 (3 and 4) are the topological interface. A hinge is formed in the center of the hinge supercell. (d) Band diagram of the hinge supercell. The hinge states, surface states, and bulk states are denoted by the purple, orange, and black lines, respectively. The inset illustrates the sound pressure field around the hinge for $k_y = 0$. The eigenstates around the hinge at $k_y = 0$ (frequency 11.88 kHz) are illustrated in the inset. (e) The fabricated sample and experiment setup. (f) The simulated and measured sound pressure field on the scan area at 11.40 kHz.

To confirm this scenario, we use two supercells S2 ($h_1 = 0.175a$, $h_2$ for SC1 $= 0.45a$) and S3 ($h_1 = 0.175a$, $h_2$ for SC2 $= 0.45a$), as introduced in Fig. 2(a), to form such a domain wall. The structures of S2 and S3 are illustrated in Supplemental Material Section 1. Their acoustic band structures are presented in Figs. 4(a) and 4(b), respectively. For S2, the surface band gap is trivial, whereas the surface band gap for S3 is topological.

We build a slab-like supercell consisting of 6 S2 supercells and 6 S3 supercells to reveal the hinge states [see Fig. 4(c)]. We denote the two sonic crystals comprising the S2 supercell as 1 and 2, and the two sonic crystals comprising the S3 supercell as 3 and 4. The slab-like supercell is periodic in the $y$ direction, but finite in both the $x$ and $z$ directions. A 1D hinge boundary along the $y$ direction is shared by the 1, 2, 3, and 4 regions. Due to the distinct interface states at the (1, 2) and (3, 4) interfaces, topological boundary states emerge at the 1D hinge. As shown by the acoustic band structure in Fig. 4(d), there are surface states and hinge states in the bulk band gap. In particular, the hinge states emerge in the surface band gap, confirming the above picture of surface topology driven hinge states.

To verify these results in experiments, we fabricate a sample consisting of 8 layers of the slab-like supercell [see Fig. 4(e)]. A speaker is used to generate sound waves impinging on the left side of the sample through a square tube (side length $5a$). The sound pressure field at the $x$-$z$ plane 1 mm away from the right surface is scanned. At the excitation frequency of 11.40 kHz, the simulated and measured acoustic pressure profiles are shown in Fig. 4(f). Both acoustic pressure profiles consistently show the emergence of hinge localized states. At frequencies higher or lower (such as 13.30 and 10.80 kHz), sound waves propagate through the sample two-dimensionally via the topological surface states of supercell S2 or S3 (see the experimental results in Supplemental Material Section 10). These results confirm the emergence of the topological hinge states due to the distinct topology of the positive and negative Dirac mass surface states.

## 5. Conclusion

In this research, we design specifically a 3D acoustic higher-order TI with three topological surface bands. The 2D topological surface states can be described by the spin-1 Dirac equation. The surface bands and the corresponding Dirac mass can be tuned by changing the geometry of the sonic resonators. Accompanying the closing and reopening of surface band gap, there is a sign reversal of the Dirac mass and hence the topological transition. At the transition point, the surface Dirac physics emerges in the bulk band gap, and the surface states become 2D massless Dirac waves. The simultaneous zero effective mass density and infinite bulk modulus at the Dirac point are experimentally validated. Topological hinge states emerge when the 2D Dirac mass of the gapped surface states changes from positive to negative. 1D sound wave propagation via the topological hinge states and the dimensional change of sound waves are realized and experimentally verified. Our findings reveal the intriguing Dirac physics in higher-order topological materials, which can lead to practical applications to manipulate acoustic waves.

**Acknowledgments**

F. M is supported by the National Natural Science Foundation of China (Grant No. 12102315), Z.-K. L and J.-H. J are supported by the National Natural Science Foundations of China (Grant No. 12074281) and the Jiangsu Distinguished Professor Funding. W. L, X H and B. J are

supported by the Australia Research Council (Grants Nos. DP190103186, DP210103523, FT210100806).

*Meng, et al. Supplemental Material*

## 1. Structure design of the unit cells and supercells

The two sonic crystals include a topologically non-trivial sonic crystal SC1 and a topologically trivial sonic crystal SC2. SC1 is firstly constructed by some connected air cavities, which are respectively located at each face of a cube and coupled through their overlapped regions. Then SC2 is realized through shifting the unit cell of SC1 by a vector of $\Delta r = (0.5a, 0.5a, 0.5a)$ and shows different topological properties from SC1.

The shape of the air cavities at each face is designed as the superposition of 2 pyramid structures, labelled as pyramid 1 and pyramid 2. Their positions are illustrated in Fig. 1a in the main text. Pyramid 1 (2) has a base length of $b_1$ ($b_2$) and a height of $h_1$ ($h_2$). In this research, $b_1$ and $b_2$ are taken as $0.955a$ and $0.675a$, respectively, while $h_1$ and $h_2$ vary. Note that if $h_1 \geq h_2$, pyramid 1 covers pyramid 2 completely.

When $h_1 = h_2 = 0.175a$, we can obtain SC1 and SC2 depicted in Fig.1b in the main text, and the supercell S1 can be composed from them following the process in Fig. S1. By increasing $h_2$ for SC1 or SC2 to $0.45a$, we can obtain the modified sonic crystals SC1* or SC2* with bigger air cavities. Please note that their topological properties remain unchanged. Then, as displayed in Fig. S2, combining SC1* with SC2, we can obtain the supercell S2; combining SC2* with SC1, we obtain the supercell S3.

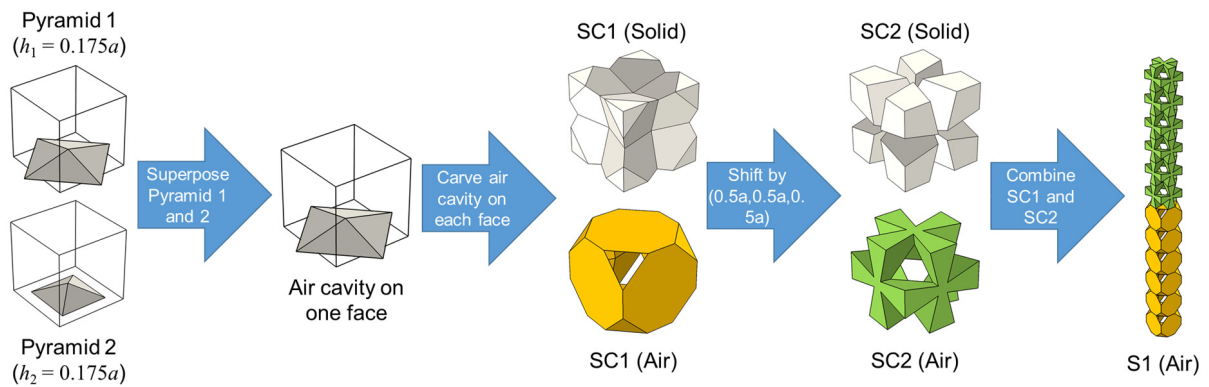

Fig. S1 | Structure design of the unit cells and supercell S1.

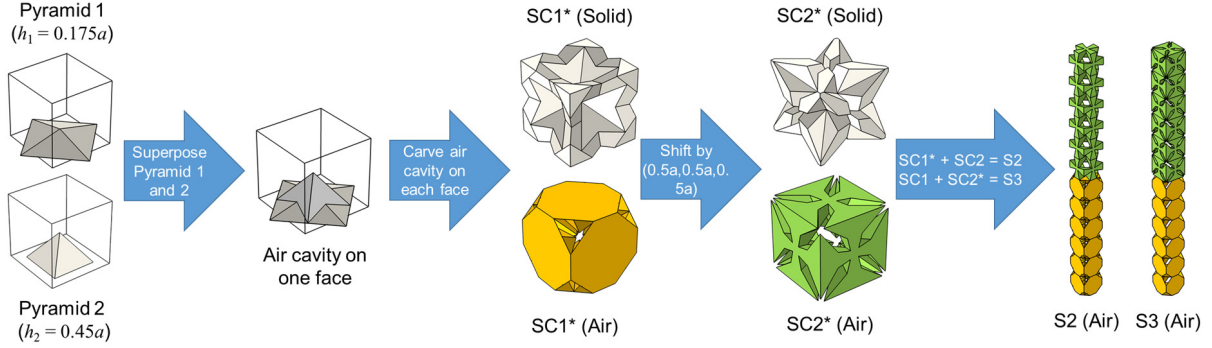

Fig. S2 | Structure design of the unit cells and supercells S2 and S3.

## 2. Fractional bulk polarization

For the two sonic crystals, Berry curvature vanishes everywhere in the first Brillouin zone (BZ) [1,2] due to the simultaneous restrictions of inversion symmetry of the simple cubic lattice and the time-reversal symmetry. The topological properties of these sonic crystals can instead be characterized by the fractional bulk polarization, $\mathbf{P}$ $(P_x, P_y, P_z)$. It is the integration of the Berry connection over the momentum space [1,3],

$$\mathbf{P} = -\frac{1}{(2\pi)^3} \iiint dk_x dk_y dk_z \mathrm{Tr}[\mathbf{A}_n(\mathbf{k})], \tag{S1}$$

where $\mathbf{A}_n(\mathbf{k}) = i\langle u_n(\mathbf{k})|\partial_\mathbf{k}|u_n(\mathbf{k})\rangle$ is the Berry connection, $\mathbf{k} = (k_x, k_y, k_z)$ is the wavevector. $n$ refers to the band index which runs over all the bands below the band gap. $\partial_\mathbf{k}$ is the vector gradient operator in $\mathbf{k}$-space. $|u_n(\mathbf{k})\rangle$ is the periodic part of the Bloch wave function. The integration in Eq. S1 is conducted over the first BZ. Due to the crystalline symmetry of the space group $P_{m\bar{3}m}$ (including three mirror symmetries and threefold rotation symmetry along [1,1,1]), we have $P_x = P_y = P_z$, and $P_i$ is quantized to either $0$ or $1/2$. The quantized value can be determined by the parities of the Bloch eigenstates at the high-symmetry points (HSPs) in the BZ [1,3]:

$$P_i = \frac{1}{2}(\sum_n q_i^n \bmod 2), \ (-1)^{q_i^n} = \frac{\eta_n(\mathrm{X}_i)}{\eta_n(\Gamma)}, \tag{S2}$$

The summation is taken over all the bands below the band gap. $i = x, y, z$ represents the direction. $\eta_n(\mathrm{X}_i)$ is the parity of the $n^{\mathrm{th}}$ band at the HSPs X ($\mathbf{k} = (a/\pi, 0, 0)$), Y ($\mathbf{k} = (0, a/\pi, 0)$), and Z ($\mathbf{k} = (0, 0, a/\pi)$). For the sonic crystals SC1 and SC2, the sound pressure profiles of the eigenstates at HSPs Γ and X and their corresponding parities are illustrated in Fig. S3~S6, where "+" indicates the even parity while "−" the odd parity. Note that the parities of Y and Z are the same as X due to the mirror symmetries of the simple cubic lattice. From Eq. S2 we can get $\mathbf{P} = (1/2, 1/2, 1/2)$ for SC1, which indicates a topological phase, while $\mathbf{P} = (0, 0, 0)$ for SC2, denoting a trivial phase. The topology here is protected by the crystalline symmetry of $P_{m\bar{3}m}$ space group [4].

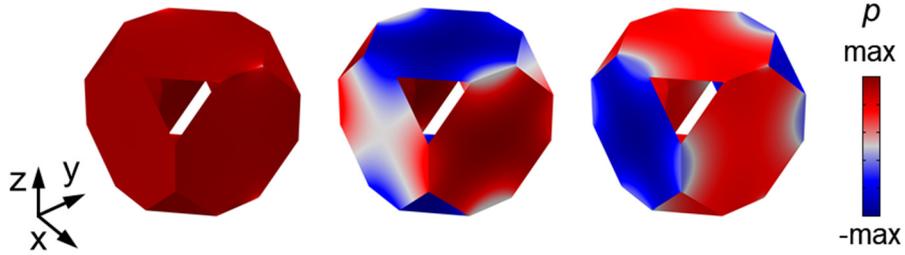

Fig. S3 | Eigenstates at Γ point for bands 1, 2, 3 of SC1. The parities are +, +, +, respectively.

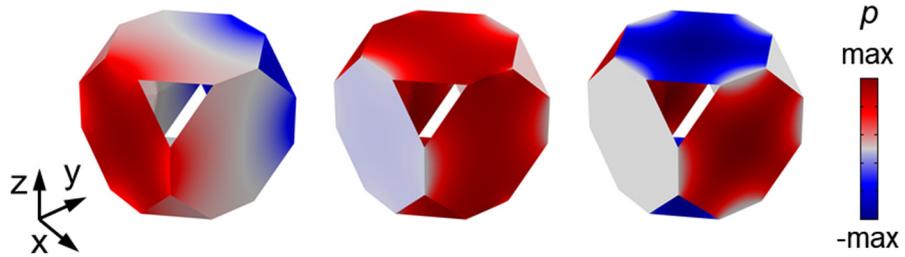

Fig. S4 | Eigenstates at X point for bands 1, 2, 3 of SC1. The parities are +, −, +, respectively.

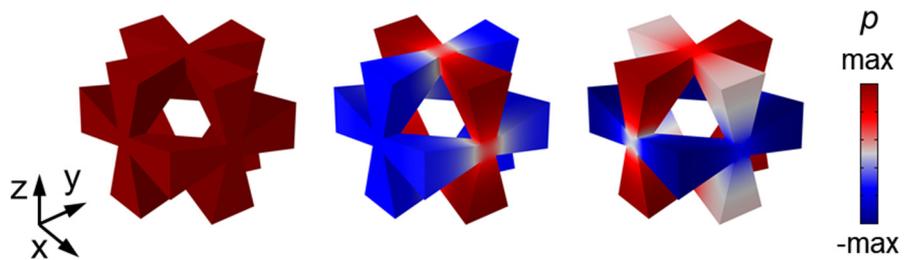

Fig. S5 | Eigenstates at Γ point for bands 1, 2, 3 of SC2. The parities are +, +, +, respectively.

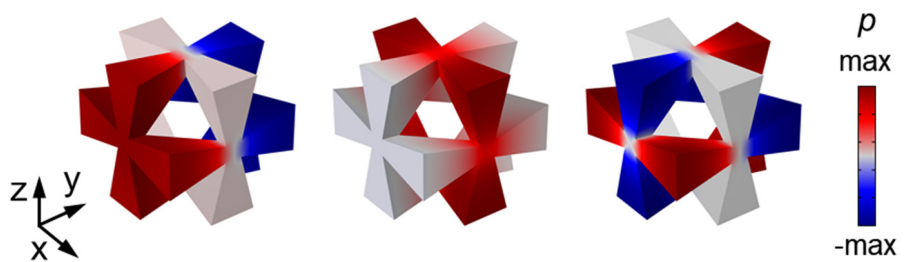

Fig. S6 | Eigenstates at X point for bands 1, 2, 3 of SC2. The parities are −, +, −, respectively.

## 3. Topological index of interface states

In this work, the interface band inversion occurs with the closing and reopening of the interface band gap, indicating that we can consider only the interface bands to characterize the topological transition. Following Ref. [5], as the supercell with interface bands enjoys $C_4$ point group symmetry, the topology of the interface band gap can be detected by the following topological index

$$\chi = ([\bar{X}_1^{(2)}], [\bar{M}_1^{(4)}], [\bar{M}_2^{(4)}]), \tag{S5}$$

where $[\Pi_p^{(n)}] = \#\Pi_p^{(n)} - \#\bar{\Gamma}_p^{(n)}$, $\#\Pi_p^{(n)}$ is the number of acoustic interface bands below the interface band gap with $C_n$ symmetry eigenvalue $e^{ip(2\pi/n)}$ at $\Pi$ point, $p = 0, 1, \ldots, n-1$ and $\Pi$ stands for $\bar{X}$ and $\bar{M}$.

In our case, however, the first interface state at the $\bar{\Gamma}$ point always merges into and hybridizes with the bulk bands, which leaves the symmetry representations intangible. To solve the problem, we instead turn to the real-space descriptions of the acoustic interface bands and then infer the hidden symmetry representations in momentum space. Firstly, based on the symmetry representations of the acoustic bulk states below the bulk band gap at $\Gamma$, $X$, $M$ and $R$ points, we search the elementary band representations (EBRs) of the space group $P_{m\bar{3}m}$ and conclude the band representations as $(A_{1g} \uparrow G)_{3d}$ and $(A_{1g} \uparrow G)_{3c}$ for SC1 and SC2, respectively [6]. In other words, the acoustic bulk states are described by s Wannier orbitals located at three inequivalent Wyckoff positions $3d$ (center of the interface) for SC1 and Wyckoff positions $3c$ (center of hinges) for SC2, as depicted in Figs. S7a and S7b. The three-dimensional bulk polarization $(1/2, 1/2, 1/2)$ of SC1 and $(0, 0, 0)$ of SC2 can also be inferred from the Wannier orbital distribution.

Considering that the supercell with interface states forms an interface of SC1 and SC2, the Wannier orbitals locate at the interface can be easily obtained as $(A \uparrow G)_{1a} \oplus (A \uparrow G)_{2c}$ (see Fig. S7c), which are physically related to the interface states. As given in Table S1a, these Wannier orbitals induce the symmetry representations in momentum space as $A_{\bar{\Gamma}} + A_{\bar{X}} + A_{\bar{M}}$ and $(A \oplus B)_{\bar{\Gamma}} + (A \oplus B)_{\bar{X}} + (1_E 2_E)_{\bar{M}}$, forming the single interface band and two degenerate interface bands at the $\bar{M}$ point, respectively. Therefore, the interface bands below the interface band gaps for S2 and S3 separately correspond to these two band representations. Based on the character tables of $C_2$ and $C_4$ symmetry (see tables S1b and S1c), we can acquire the $C_n$ symmetry eigenvalues of these representations at $\bar{\Gamma}$, $\bar{X}$, and $\bar{M}$ points. Then, the topological indexes of the interface band gaps for S2 and S3 are calculated as $\chi = (0,0,0)$ and $\chi = (-1,-1,1)$, respectively. The different topological indexes imply the topological transition among the interface states. Furthermore, the two-dimensional polarization $\mathbf{P}$ can also be deduced from the topological index $\chi$, i.e.,

$$P_x = P_y = \frac{1}{2}\left[X_1^{(2)}\right] \ mod \ 1. \tag{S6}$$

Therefore, the polarization $\mathbf{P}$ of the surface band gap is $(0,0)$ for supercell S2, while a nontrivial quantization of $(1/2, 1/2)$ for S3. The different polarization of two interface band gaps is responsible for the resultant hinge states.

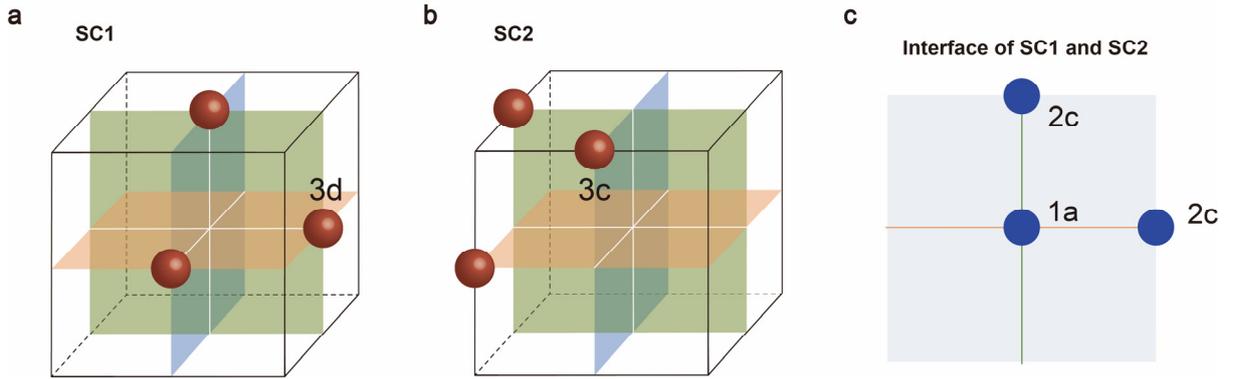

Fig. S7 | Wannier representations of acoustic bulk and interface states. (a) Wannier orbitals locate at the Wyckoff position $3d$ for SC1. (b) Wannier orbitals locate at the Wyckoff position $3c$ for SC2. (c) Wannier orbitals locate at Wyckoff positions $1a$ and $2c$ for the interface states. The red spheres and blue dots denote $s$ orbitals.

a

| EBRs | $\Gamma$ | X | M |
|---|---|---|---|
| $(s)_{1a}$ | A | A | A |
| $(s)_{2c}$ | A⊕B | A⊕B | $^1E\ ^2E$ |

b

| irreps | E | $C_2$ |
|---|---|---|
| A | 1 | 1 |
| B | 1 | -1 |

c

| irreps | E | $C_4^+$ | $C_2$ | $C_4^-$ |
|---|---|---|---|---|
| A | 1 | 1 | 1 | 1 |
| B | 1 | -1 | 1 | -1 |
| $^1E$ | 1 | -i | -1 | i |
| $^2E$ | 1 | i | -1 | -i |

Table. S1 | (a) Elementary band representations of $(A\uparrow G)_{1a}$ and $(A\uparrow G)_{2c}$. (b) and (c) Character tables of $C_2$ and $C_4$ point group symmetries, respectively.

## 4. The $\boldsymbol{k} \cdot \boldsymbol{p}$ theory of the interface Hamiltonian

We use the $\boldsymbol{k} \cdot \boldsymbol{p}$ theory to study the band structure near the interface Dirac-like cone [S5]. The eigenvalue problem in sonic crystals is to solve the following wave equation:

$$-\nabla \cdot [\rho_r^{-1}(\boldsymbol{r}) \nabla p_{n,\boldsymbol{k}_\|}(\boldsymbol{r})] = \omega_{n,\boldsymbol{k}_\|}^2 / v^2 K_r^{-1}(\boldsymbol{r}) p_{n,\boldsymbol{k}_\|}(\boldsymbol{r}), \tag{S7}$$

where $p_{n,\boldsymbol{k}_\|}(\boldsymbol{r})$ is the Bloch wavefunction of the acoustic pressure field with the wavevector $\boldsymbol{k}_\| = (k_x, k_y)$ in the $n^{\text{th}}$ band, $\omega_{n,\boldsymbol{k}_\|}$ is the corresponding eigenfrequency. The Bloch function is normalized as $\int_{\text{u.c.}} p_{n,\boldsymbol{k}_\|}^*(\boldsymbol{r}) \frac{1}{K_r(\boldsymbol{r})} p_{n',\boldsymbol{k}_\|}(\boldsymbol{r}) = \delta_{nn'}$, with u.c. denoting the unit cell. $\rho_r(\boldsymbol{r}) = \rho(\boldsymbol{r})/\rho_0(\boldsymbol{r})$ and $K_r(\boldsymbol{r}) = K(\boldsymbol{r})/K_0(\boldsymbol{r})$ denote the relative constitutive mass density and bulk modulus, respectively, $v = \sqrt{K_0/\rho_0}$ is the speed of sound in the air host. The Hermitian operator $-\nabla \cdot [\rho_r^{-1}(\boldsymbol{r}) \nabla]$ can be viewed as the Hamiltonian of the acoustic pressure field.

The key idea of the $\boldsymbol{k} \cdot \boldsymbol{p}$ theory here is using three Bloch wavefunctions $p_{j,\boldsymbol{k}_0}$ ($j = 1, 2, 3$) at the interface Dirac-like point $\boldsymbol{k}_0$ to expand the wavefunctions at $\boldsymbol{k}$ around $\boldsymbol{k}_0$. Recalling that the Bloch wavefunctions have the form $p_{n,\boldsymbol{k}_\|}(\boldsymbol{r}) = e^{i\boldsymbol{k}_\| \cdot \boldsymbol{r}} u_{n,\boldsymbol{k}_\|}(\boldsymbol{r})$ with $u_{n,\boldsymbol{k}_\|}(\boldsymbol{r})$ being a complete basis set in Hilbert space, we can expand $p_{n,\boldsymbol{k}_\|}(\boldsymbol{r})$ around $\boldsymbol{k}_0$ as the linear combination of $p_{j,\boldsymbol{k}_0}(\boldsymbol{r})$, i.e.,

$$p_{n,\boldsymbol{k}_\|}(\boldsymbol{r}) = \Sigma_j A_{n,j}(\boldsymbol{k}_\|) e^{i(\boldsymbol{k}_\| - \boldsymbol{k}_0) \cdot \boldsymbol{r}} p_{j,\boldsymbol{k}_0}(\boldsymbol{r}), \tag{S8}$$

where $A_{n,j}$ are expansion coefficients. In principle, the band indices $j$ runs over all bands at $\boldsymbol{k}_0$. In practice, however, we are only interested in three bands in the vicinity of the Dirac-like points, the $\boldsymbol{k} \cdot \boldsymbol{p}$ theory can be restricted to the Hilbert space consisting of only three Bloch wavefunctions. Substituting Eq. S8 into Eq. S7 and utilizing the orthogonality, we obtain the following Hamiltonian:

$$H_{lj}(\Delta \boldsymbol{k}_\|) = \frac{\delta_{lj} \omega_{j,\boldsymbol{k}_0}^2}{v^2} + \boldsymbol{p}_{lj} \cdot \boldsymbol{q}, \tag{S9}$$

here, $\boldsymbol{q} = \boldsymbol{k}_\| - \boldsymbol{k}_0$, we have omitted the high-order term of $|\boldsymbol{q}|$. The $\boldsymbol{p}_{lj} \cdot \boldsymbol{q}$ term is analogous to the $\boldsymbol{k} \cdot \boldsymbol{p}$ term in the electronic problem. The matrix element $\boldsymbol{p}_{lj}$ is given by

$$\boldsymbol{p}_{lj} = -i \int_{u.c.} p_{l,\boldsymbol{k}_0}^*(\boldsymbol{r}) \left\{ \frac{2\nabla p_{j,\boldsymbol{k}_0}(\boldsymbol{r})}{\rho_r(\boldsymbol{r})} + \left[\nabla \frac{1}{\rho_r(\boldsymbol{r})}\right] p_{j,\boldsymbol{k}_0}(\boldsymbol{r}) \right\} d\boldsymbol{r}, \tag{S10}$$

which represents the mode-coupling integrals between degenerate states at the Dirac-like point and determines the linear dispersions around $\boldsymbol{k}_0$. The known is that, $\rho_r(\boldsymbol{r})$ and the operator $\nabla$ are of even and odd parities, respectively, $\boldsymbol{p}_{lj}$ is nonzero only when $p_{l,\boldsymbol{k}_0}(\boldsymbol{r})$ and $p_{j,\boldsymbol{k}_0}(\boldsymbol{r})$ are of opposite parities. The acoustic pressure profiles in Fig. 1d and Fig. 2d in the main text show that for three states at interface Dirac-like points at both the $\bar{M}$ and $\bar{\Gamma}$ points, there is one $s$-like state of even parity and two $p$-like states of odd parity. Furthermore, considering the constraints of the $C_4$ rotation symmetry, the time-reversal symmetry $\Theta$ ($\Theta = K$, $K$ is the complex conjugation), and the mirror symmetries at $\boldsymbol{q} = (q_x, 0)$ and $(0, q_y)$, we reformulate the Hamiltonian on the basis $[s, p_x, p_y]^T$ as

$$H(\boldsymbol{q}) = \begin{bmatrix} \omega_{s,\boldsymbol{k}_0}^2/v^2 & q_x a & q_y b \\ -q_x a & \omega_{p_x,\boldsymbol{k}_0}^2/v^2 & 0 \\ -q_y b & 0 & \omega_{p_y,\boldsymbol{k}_0}^2/v^2 \end{bmatrix}, \tag{S11}$$

where $|a| = |b|$, $a$ and $b$ are pure imaginary numbers. The band dispersion $\omega_{n,\boldsymbol{q}}$ around $\boldsymbol{k}_0$ can be obtained by calculating the following secular equation:

$$\det \begin{vmatrix} \frac{\omega_{n,\boldsymbol{q}}^2 - \omega_{s,\boldsymbol{k}_0}^2}{v^2} & -q_x a & -q_y b \\ q_x a & \frac{\omega_{n,\boldsymbol{q}}^2 - \omega_{p_x,\boldsymbol{k}_0}^2}{v^2} & 0 \\ q_y b & 0 & \frac{\omega_{n,\boldsymbol{q}}^2 - \omega_{p_y,\boldsymbol{k}_0}^2}{v^2} \end{vmatrix} = 0, \tag{S12}$$

which gives two linear bands intersecting with a flat band when $\omega_{s,\boldsymbol{k}_0}^2 = \omega_{p,\boldsymbol{k}_0}^2$ (see Fig. S8), resembling the dispersion in Fig. 2c in the main text.

We then define a mass term $m = \omega_{p,\boldsymbol{k}_0}^2 - \omega_{s,\boldsymbol{k}_0}^2/v^2$, the interface Dirac point is gapped when $m \neq 0$, as shown in Fig. S8. Specifically, the interface band gaps with positive and negative $m$ are topologically distinct. In sonic crystals, $m$ is controlled by the geometric parameters and the topology is characterized by the topological invariant $\chi$ when considering the full surface BZ. The topological phase diagram versus "$m$" is shown in Fig. 2a in the main text.

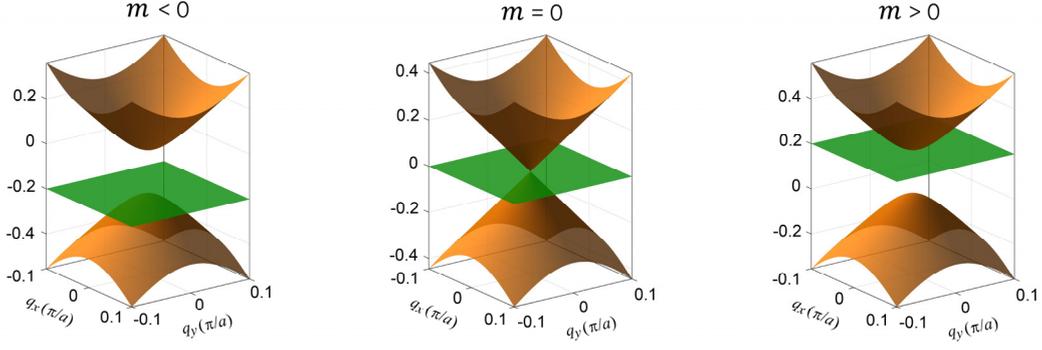

Fig. S8 | $\mathbf{k} \cdot \mathbf{p}$ interface band structures around $\mathbf{q} = 0$ with $m > 0$, $m = 0$, and $m < 0$. The Dirac point where two linear bands intersect with a flat band emerges when $m$ vanishes. Band gaps with different topologies are open when $m \neq 0$. The parameters in the $\mathbf{k} \cdot \mathbf{p}$ Hamiltonian are chosen as $a = b = i$ and $m = \pm 0.2$ (we assume they are dimensionless).

## 5. Tight-binding model of the interface states

Based on the Wannier description where three single $s$-wave orbitals are separately occupied at the sites A, B and C (see Fig. S9a), a full Lieb lattice tight-binding (TB) Hamiltonian can be constructed to describe the intriguing physics at the 2D interface. We label three $s$ orbitals as $\phi_A$, $\phi_B$ and $\phi_C$, then, the Hamiltonian matrix elements are defined via

$$H_{ij}(\mathbf{R}) = \langle \phi_{0i} H \phi_{\mathbf{R}j} \rangle, \tag{S13}$$

where $\mathbf{R}$ labels the relative lattice vector, $i$ and $j$ denote the sites A, B and C. The matrix element corresponds to a hopping from the orbital $\phi_j$ in the reference cell to the orbital $\phi_i$ in the cell $\mathbf{R}$. Using the Fourier transform that $|\chi_j^{\mathbf{k}}\rangle = \sum_{\mathbf{R}} e^{i\mathbf{k} \cdot (\mathbf{R}+\tau_j)} |\phi_{\mathbf{R}j}\rangle$ with $\tau_j$ being the relative coordinate vector of the orbitals, we obtain

$$H_{ij}^{\mathbf{k}} = \langle \chi_i^{\mathbf{k}} | H | \chi_j^{\mathbf{k}} \rangle = \sum_{\mathbf{R}} e^{i\mathbf{k} \cdot (\mathbf{R}+\tau_j-\tau_i)} H_{ij}(\mathbf{R}). \tag{S14}$$

We set the couplings between the orbitals at A and B (C) as $t_1$ ($t_2$), also, $\epsilon_A$, $\epsilon_B$ and $\epsilon_C$ as the onsite energy of three orbitals themself. Then, a $3 \times 3$ TB Bloch Hamiltonian can be written as

$$H(\mathbf{k}) = \begin{pmatrix} \epsilon_A & 2t_1 \cos\left(\frac{k_x}{2}\right) & 2t_1 \cos\left(\frac{k_y}{2}\right) \\ & \epsilon_B & 2t_2 \cos\left(\frac{k_x}{2} - \frac{k_y}{2}\right) + 2t_2 \cos\left(\frac{k_x}{2} + \frac{k_y}{2}\right) \\ h.c. & & \epsilon_C \end{pmatrix}, \tag{S15}$$

where the lattice constant is set to unity. The corresponding band structures are shown in Figs.

S9b-d, as we can see, the difference of the onsite energy at A and B (C) can play the role of the mass term and contribute to the topological phase transition alike to that of the interface states.

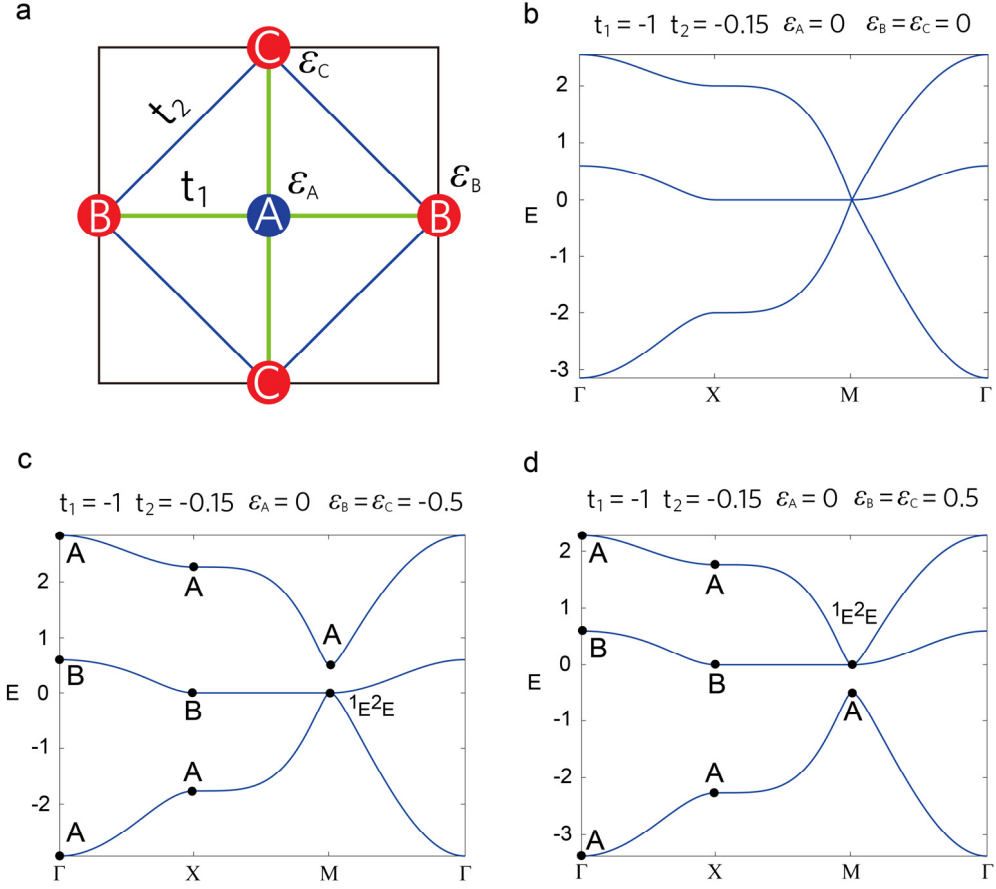

Fig. S9 | Tight-binding model of the interface states. (a) Illustration of the unit cell of the Lieb lattice which captures the physics of the interface. $t_1$ ($t_2$) denotes the coupling between the sites A and B (C). $\epsilon_A$, $\epsilon_B$ and $\epsilon_C$ are onsite energy of three sites. (b)-(d) Band structures of the Lieb lattice with different onsite energy which show the same phase transition as that of the interface states. The symmetry representations at high symmetry points are given to demonstrate the topology. The parameters are given above each figure.

## 6. Band folding of supercell

When $h_1 = h_2 = 0.27a$ for SC1 and SC2, the resultant supercell is shown in Fig. S10a. For this supercell, the MO mode and the two DI modes accidentally degenerate at the $\bar{M}$ point of the 2D surface BZ. In order to demonstrate the property of the accidental degenerated Dirac-like point, we employ the band folding mechanism and map the Dirac-like point to the $\bar{\Gamma}$ point. The expanded supercell is shown in Fig. S10b. The BZs before and after the band folding are shown in Fig. S10c.

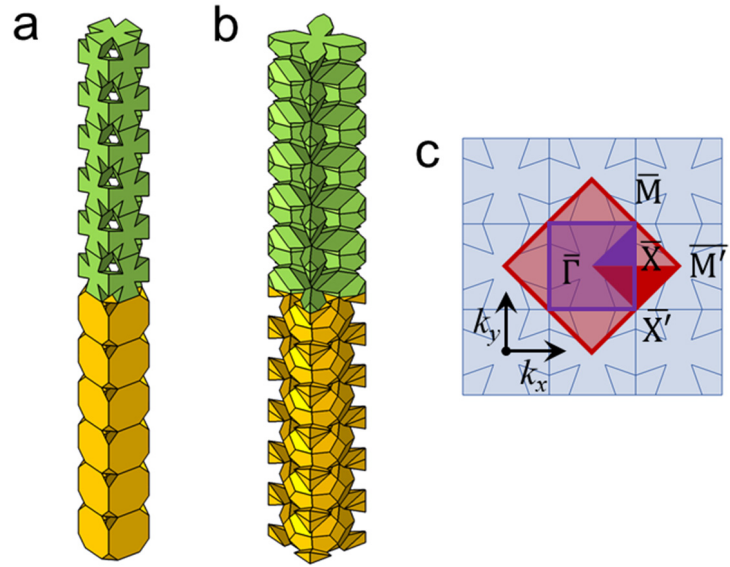

Fig. S10 | Band folding mechanism and the expanded supercell. (a) The supercell when $h_1 = h_2 = 0.27a$. (b) The expanded supercell after band folding. (c) Folding and mapping of BZ. The purple and red areas are the first BZs of the original and expanded supercells, respectively.

## 7. Gapless topological surface state

The numerically calculated band diagram reveals that the expanded supercell in main text Fig. 2 has gapless topological surface states. To validate that they are gapless, we conduct a simple experiment using the same setup in main text Fig. 3. The acrylic slab with a hole is still glued on the left side of the sample and introduces a point-like sound source. A microphone is located in the middle of the right side of the sample, close to the interface between the trivial and non-trivial sonic crystals. The sound signal in the frequency range 10 kHz to 18 kHz is picked up.

The normalized sound pressure is illustrated in Fig. S11(b). The bulk band gap of the expanded supercell, in the frequency range 12.77 to 15.51 kHz, is illustrated by color grey. It can be seen that remarkable sound propagation is detected in this frequency range, with no noticeable drop or gap. The Dirac-like surface state of the expanded supercell is indeed gapless.

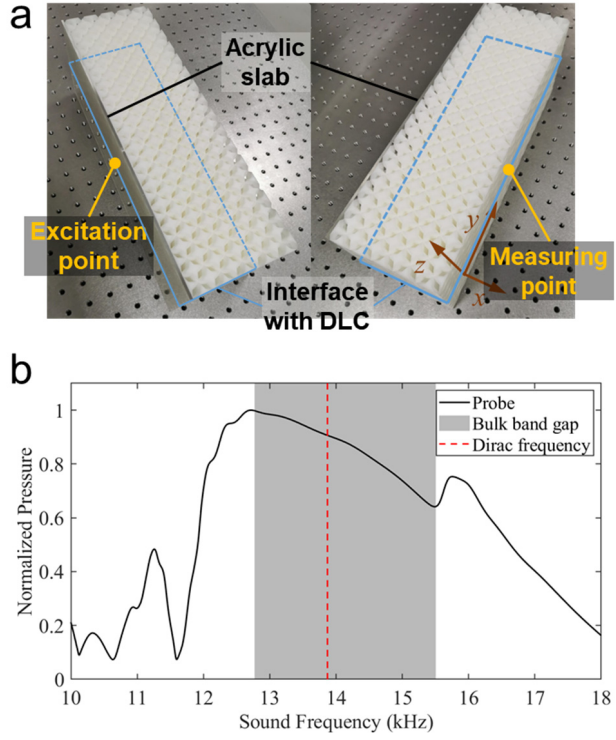

Fig. S11 | Detection of the gapless Dirac-like surface state. (a) Experiment setup. (b) Test result.

## 8. Effective parameter around the Dirac-like point

We use the effective medium theory to investigate the wave propagation properties of the Dirac-like conical interface states. The interface can be regarded as a 2D material. Using the effective parameter retrieving method from Ref. [7], we obtain the normalized effective density and the inverse of bulk modulus for frequencies around the Dirac-like point, as depicted in Fig. 2e in the main text. Fig. S12 illustrates the supercell for the effective parameter retrieval method. Assuming only one eigenstate is excited when an interface wave impinging on the left of the supercell. The impedance $Z$ of this 2D material is defined as:

$$Z_x = \frac{\langle P_x \rangle}{\langle v_x \rangle}, \tag{S3}$$

where $P$ is the acoustic pressure field on the left face around the interface. $v$ is the acoustic velocity in the direction perpendicular to the left faces. $\langle ... \rangle$ represents the average of $P$ and $v$. Considering the symmetry of the supercell, we conclude $Z_y = Z_x$.

Assuming the supercell is a uniform 2D medium, the effective mass density $\rho$ and the bulk modulus $K$ can be obtained as:

$$\rho = \frac{k_x \cdot Z_x}{\omega}, \quad K = \frac{k_x}{\omega \cdot Z_x}, \tag{S4}$$

by varying $k_x$, we can obtain the relationship between $\rho/\rho_0$, $K_0/K$ and frequencies around the Dirac-like point.

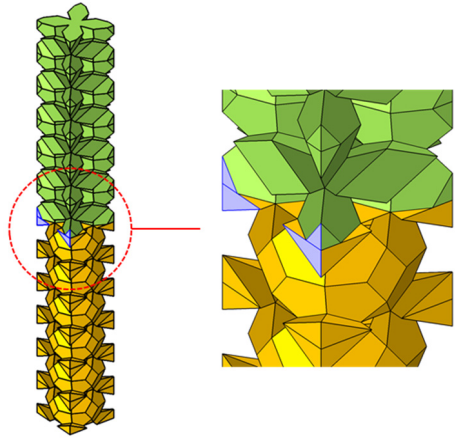

Fig. S12 | Supercell for calculating the effective parameters of the Dirac-like interface states. The pressure field and the velocity field on the blue faces in the vicinity of the interface are considered in the calculation.

## 9. Immunity of hinge states to imperfections

One of the remarkable features of TI is that the topological edge states are immune to imperfections. Here, to validate the robustness of the hinge states, we deliberately introduce a substantial number of defects into the hinge supercell. As shown in Fig. S13a, for instance, 8 unit cells of the supercell are removed. These defects are $2a$ away from the hinge. For the defective hinge supercell, the hinge states still exist and the corresponding eigenfrequencies (denoted by the purple dots in Fig. S13b) are almost unaffected. The average relative error of the frequencies is about 0.20%, which endorses the robustness of the hinge state. The sound pressure profile of the defective hinge supercell at $k_y = 0$ is depicted in Fig. S13c. The acoustic wavefunction highly concentrates on the hinge, which is the same as the sound pressure profile in Fig. 4d.

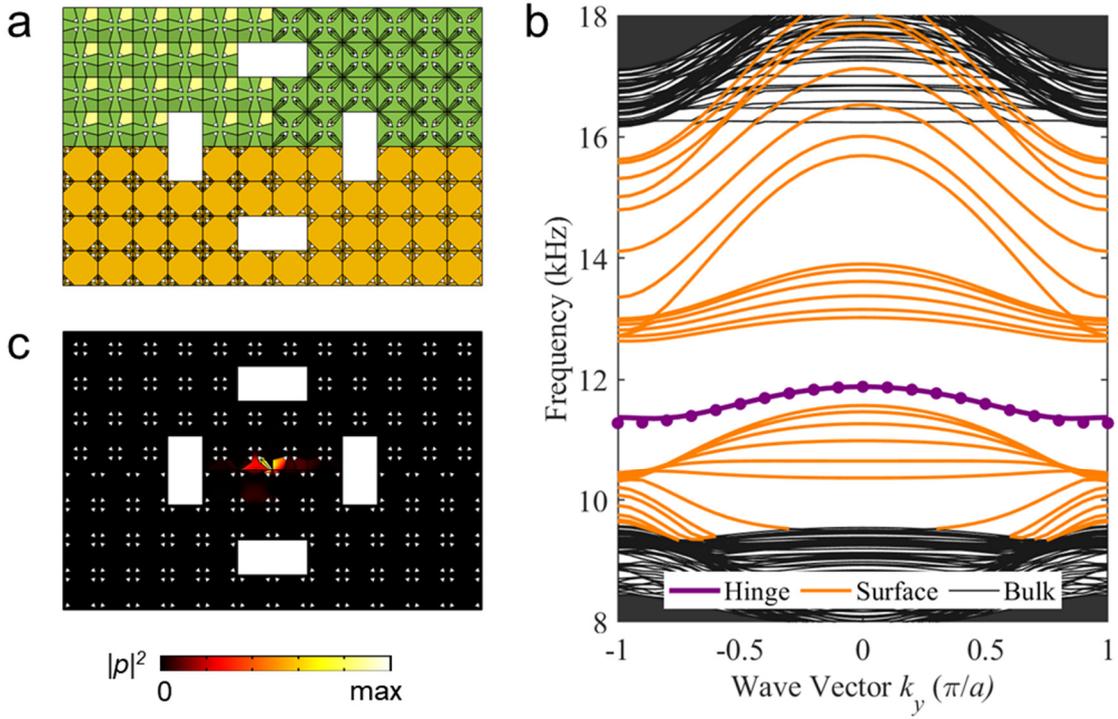

Fig. S13 | Robustness of 1D hinge states. (a) The defective hinge supercell after removing 8 unit cells around the center. (b) Band diagram of the defective hinge supercell, where the hinge states are denoted by the purple line and dots. (c) Sound pressure field of the hinge state at $k_y$ = 0 after introducing defects. The acoustic wavefunction remains localized at the center of the supercell.

## 10. Multidimensional propagation of sound in the hinge model

The hinge state in Fig. 4(d) is within a narrow frequency range 11.37 kHz to 11.88 kHz. Using the experiment in the main text [Fig. 4(f)], we have visualized the hinge state. By choosing more frequencies, we can investigate the dimensional evolution of sound propagation. The sound pressure field on the scan plane for two other frequencies, 13.30 and 10.80 kHz, are simulated and measured. They are illustrated in Fig. S14 together with the sound profile for 11.40 kHz. When the frequency change from 13.30 to 11.40 and then to 10.80 kHz, the acoustic energy firstly localizes on the 2D topological surface of S2, then localizes on the 1D hinge, and finally change to the 2D topological surface of S3. The experiment results agree well with the theoretical prediction, visualize the existence of hinge state and the dimensional hierarchy of propagation modes.

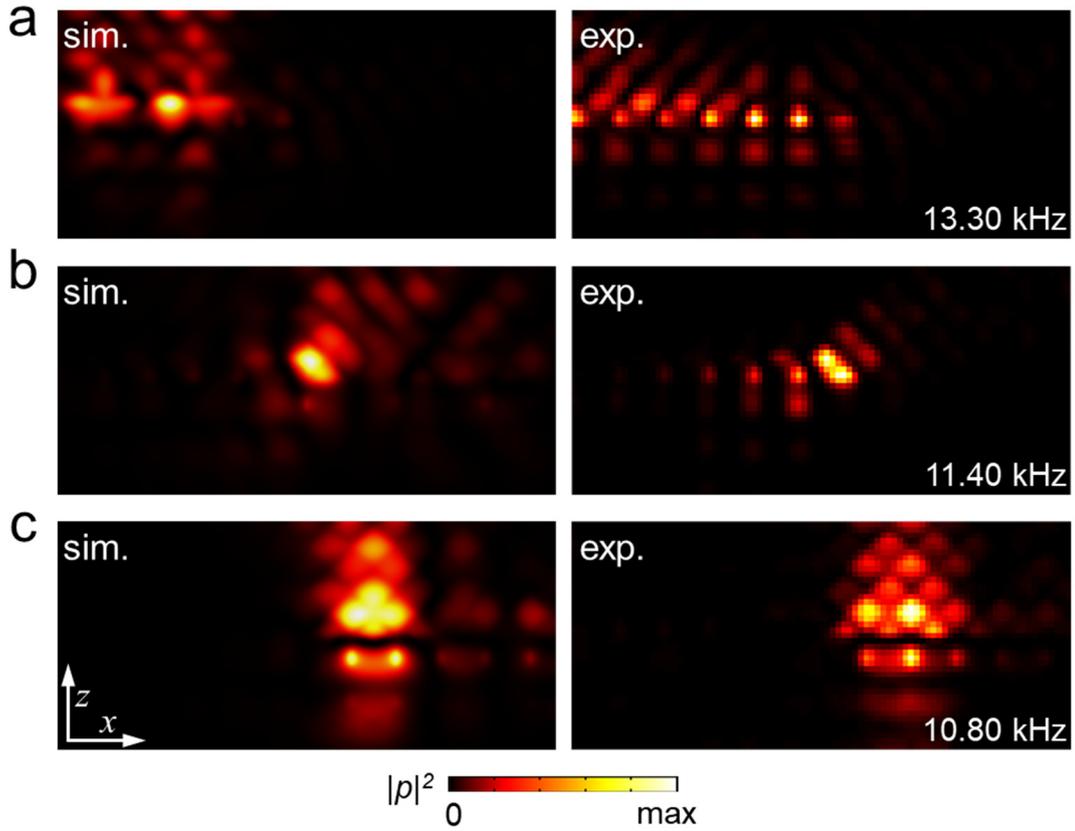

Fig. S14 | Dimensional evolution of sound propagation. (a) (b) (c) The simulated (left) and measured (right) sound pressure field on the scan area for 13.30, 11.40, and 10.80 kHz respectively. With the decreasing of the frequency, the results denote the evolution of the sound transportation from the topological surface states of S2 (13.30 kHz), to the hinge state (11.40 kHz), and then to the topological surface states of S3 (10.80 kHz).

## 11. Methods

Numerical simulations in this work are conducted by the pressure acoustic module of COMSOL Multiphysics. The density of air is taken as 1.21 kg·m$^{-3}$, sound speed in air is 343 m·s$^{-1}$, bulk modulus of air is $1.42\times10^5$ Pa. The sample is made of photosensitive resin via stereolithography (SLA), with a fabrication error of 0.1mm. The air-structure interfaces are treated as sound hard boundaries considering acoustically rigid materials construct the sonic crystals. The sound source is a HIVI RT1C-A speaker. The sine wave sound signal is generated by the built-in sound card of BSWA MC3242 data collector. Sound pressure is picked up by NI 9233 data acquisition card with BSWA MPA416 microphones. The sound profiles are picked up by a microphone fixed on a motorized linear stage. The step length between two scan points is 2 mm.